\documentstyle[12pt]{article}
\begin{document}
\hfill{LANDAU-95-GR-2}

\vspace{0.5cm}

\centerline{\bf SPECTRUM OF INITIAL PERTURBATIONS}
\centerline{\bf IN OPEN AND CLOSED INFLATIONARY MODELS}
\bigskip
\centerline{A.A. Starobinsky}
\centerline{Landau Institute for Theoretical Physics,}
\centerline{Russian Academy of Science, Moscow 117334, Russia}

\vspace{0.5cm}
\noindent
{\it Talk presented at the First International Conference on Cosmoparticle
Physics "Cosmion-94" (Moscow, 5-14 December 1994). Published in:
Cosmoparticle Physics. 1, eds. M.Yu. Khlopov, M.E. Prokhorov, A.A.
Starobinsky, J. Tran Thanh Van, Edition Frontiers, 1996, pp. 43-52.}

\vspace{0.5cm}
\noindent
{\bf Abstract.} 
Spectrum of initial scalar and tensor
perturbations created during an inflationary stage producing a closed or
open FRW universe now is discussed. In the closed case, the CMB temperature
anisotropy ${\Delta T/T}$ generated by scalar perturbations is enhanced for
low multipoles. It is argued that in the open case there is no suppression
of low multipoles. A possibility of the existence of a
preferred space direction in the open case is noted.

\vspace{1cm}

{\centerline {\bf 1. Introduction.}}

\vspace{0.5cm}

Though the prediction that total present energy density of matter in the
Universe (including the cosmological constant if it is non-zero) should be
equal to the critical one $\varepsilon_c=3H_0^2/8\pi G$ ($H_0$ is the Hubble
constant, and we assume $c=\hbar=1$) is usually considered as one of the
basic predictions of the inflationary scenario of the early Universe, it is
not an absolute prediction. More complicated inflationary models can be
constructed which contain at least two parameters in effective Lagrangians
describing the de Sitter (inflationary) stage and which may lead to
$\Omega_m\not= 1$ at present (thus, they belong to the second and higher
complexity levels according to the classification of cosmological models
presented in [1]). Then the first of the parameters determines an amplitude
of the approximately flat ($n_s\approx 1$) spectrum of initial adiabatic
perturbations while the second one gives the present value of $\Omega_m=1+
K/H_0^2a_0^2$. Here $a_0$ is the present value of the scale factor of the
Friedman-Robertson-Walker (FRW) cosmological model and $K=1, 0, -1$ denotes
closed, flat and open FRW models respectively.

This freedom is due to the fact that the exact de Sitter
space-time which serves as a basic element of the inflationary scenario can
be covered by different charts (systems of reference). In particular, its
metric can be represented as a partial case of all three FRW models:

\begin{eqnarray}
ds^2=dt^2_+-H_1^{-2}\cosh^2H_1t_+(d\chi^2_++\sin^2\chi_+
d\Omega^2),~~~K=+1, \\
ds^2=dt^2-a_1^2e^{2H_1t}(dr^2+r^2d\Omega^2),\,~~~~a_1=const,~~~~~K=0,  \\
ds^2=dt^2_--H_1^{-2}\sinh^2H_1t_-(d\chi^2_-+\sinh^2\chi_-
d\Omega^2),~~K=-1, \\
d\Omega^2=d\theta^2+\sin^2\theta d\varphi^2,~~~H_1^2=
\frac{\Lambda_1}{3} =const.  \nonumber
\end{eqnarray}

Note that the first metric ($K=+1$) covers the whole de Sitter space-time,
and the two others have horisons at $t=-\infty$ and $t_-=0$ respectively.
Coordinate transformations between these 3 systems of refernce are given by
the following formulas:

a) from (2) to (1)
\begin{eqnarray}
t&=&H_1^{-1}\ln(\cosh H_1t_+\cos \chi_++\sinh H_1t_+), \nonumber \\
r&=&(a_1H_1)^{-1}\frac{\cosh H_1t_+\sin\chi_+}{\cosh H_1t_+\cos\chi_++\sinh
H_1t_+};
\end{eqnarray}

b) from (3) to (1)
\begin{eqnarray}
&\cosh H_1t_-=\cosh H_1t_c \cosh H_1t_+\cos\chi_+-\sinh H_1t_c
\sinh H_1t_+,  \nonumber \\
&\sinh H_1t_-\sinh\chi_-=\cosh H_1t_+\sin\chi_+
\end{eqnarray}
where $t_c$ is an arbitrary constant. The metric (3) covers the future light
cone of the point $t_+=t_c$, $\chi_+=0$ in the metric (1). Thus, it represents
an analog of the Milne metric in the flat space-time. In the inflationary
scenario, the de Sitter space-time is neigher exact,
nor stable; it is only approximate and metastable. So, by adding homogeneous
perturbations to the metrics (1-3) which drive exact solutions away from the
de Sitter stage one can obtain a FRW Universe with $\Omega_m\not= 1$ even
in the simplest models (see, e.g., Eq.(10) of the paper [2] where the first
explicit cosmological model with the initial de Sitter stage was constructed).

However, in that case it requires special fine tuning of initial conditions
(namely, of the amplitude of these homogeneous perturbations) at the beginning
of inflation to have $\Omega_m$ significantly different from $1$ at present.
In particular, the number of e-folds during inflation should be fine-tuned to
about 70. This contradicts the spirit of the inflationary scenario. For natural
initial conditions without fine-tuning, the simplest inflationary models
predict $\Omega_m\approx 1$ with a high degree of accuracy. Actually, in this
case the deviation of $\Omega_m$ inside the present cosmological horison from
unity is determined by inhomogeneous fluctuations of the quasi-Newtonian
gravitational potential inside the horison, too, and not by an isotropic
homogeneous part of spatial curvature which is exponentially small. It can be
shown that $\Omega_m-1$ is a Gaussian stochastic quantity with the dispersion
\begin{equation}
<(\Omega_m-1)^2>=\frac{1}{8}(\xi_{\Phi}(0)-\xi_{\Phi}(2R_h))
\end{equation}
where $\xi_{\Phi}$ is the potential-potential correlation
function and $R_h$ is the horison scale. Numerically, $|\Omega_m-1|_{rms}
\approx 4(\Delta T/T)_Q \approx 3\cdot 10^{-5}$ where $(\Delta T/T)_Q\approx
7\cdot 10^{-6}$ is the expected value of the quadrupole anisotropy of
the cosmic microwave background (CMB) for $n_s=1$.

To avoid undesirable fine-tuning of initial conditions, one has eigher
to introduce an additional parameter into the inflaton potential or to add
a second inflaton scalar field (that leads to double inflation) in order to
have $70$ e-folds of inflation (or, of the last phase of inflation in the case
of double inflation) for typical initial conditions.
According to the philosophy used in [1], then the word "fine-tuning" is
no more adequate, but one should refer the resulting inflationary
model to the second level of complexity.

\vspace{0.5cm}
\centerline{\bf 2. Closed FRW universes}

\vspace{0.5cm}

This case is closely related to the hypothesis of the Universe's
"creation from nothing" [3-6].
Though this hypothesis still remains unproved (in particular, in no model
was the probability of this process calculated or at least rigorously
shown to be non-zero), it is assumed usually that this creation
should be described by an instanton (Euclidean) solution of
classical equations. An $O(3)$ instanton ($O(4)$, if the
energy-momentum tensor of the inflaton field may be approximated
by a cosmological constant) just leads to a closed FRW universe
in the Lorentzian, or Minkowskian, region. The background equations
in the Lorentzian region have the form:
$$H^2+{1\over a^2}={8\pi G \over 3}\left( {\dot \phi^2 \over 2}+
V(\phi)\right),~~~H\equiv {\dot a \over a}, $$
\begin{equation}
\ddot \phi +3H\dot \phi +{dV \over d\phi} =0,
\end{equation}
dot means differentiation with respect to $t_{+}$.
Equations for instanton configurations in the Euclidean region
follow from here after the substitution $t_{+}=i\tau$.

Let us assume now that the tunneling ends at the point $\phi=\phi_1$
(so that $\dot\phi=\dot a=0,~\phi=\phi_1$ in both the Euclidean
and Lorentzian regions at the moment $t_{+}=\tau=0$) lying in a flat
region of the inflaton potential (i.e., satisfying the slow-roll
conditions $V_1'\equiv {dV\over d\phi}(\phi_1)\ll \sqrt{48\pi G} V_1,
~V''_1\equiv {d^2V \over d\phi^2}(\phi_1) \ll 24\pi GV_1$ where $V_1\equiv
V(\phi_1)$). Then, just after tunneling an inflationary stage begins. By
choosing the number of e-folds between $\phi=\phi_1$ and the end of inflation
\begin{equation}
\ln{a_f \over a_1}=8\pi G\int_{\phi_f}^{\phi_1} d\phi~{V(\phi)\over
V'(\phi)} \approx70
\end{equation}
($a_f$ and $\phi_f$ are the values of the scale factor and the scalar field
respectively at the end of inflation), we succeed in constructing of an
inflationary model which has $\Omega_m > 1$ at present. Thus, the value
$\phi_1$ is just the abovementioned second parameter of the inflationary
model (the first one defining the amplitude of adiabatic perturbations is
$H_1^3/V_1'$ as we shall see below). Of course, the potential $V(\phi)$ should
have some specific properties around the point $\phi_1$ to facilitate tunneling
just to this point, but we shall not discuss this point further.

At the beginning of the inflationary stage $\phi \approx \phi_1$ and
$a(t_{+})=H^{-1}_1\cosh H_1t_{+},~H_1^2=8\pi GV_1/3$.  Then, integrating the
second of Eqs. (7), we find
\begin{equation}
\dot \phi=-{V'_1\over a^3}\int_0^{t_{+}}a^3\, dt=-{V'_1\over 3H_1}
\tanh^3H_1t_{+}\left(1+ {3\over \sinh^2H_1t_{+}}\right).
\end{equation}
 For $H_1t_{+}\gg 1$, $\dot\phi$ approaches the standard slow-rolling value
($-V_1'/3H_1$). So, the subsequent evolution of the background is as in
the $K=0$ case until recent times.

Quantization of a scalar field $\psi$ in the de Sitter background
was first performed in [7], and I have nothing to add here.
For a massless field ($\psi_{;i}^{;i} =0$) in the Heisenberg representation,
$$\hat\psi=\sum\limits_{n,l,m}\left(\hat
a_{nlm}\psi_n(\eta)Q_{nlm}(\chi_{+},\theta ,\varphi)+\hat
a^{+}_{nlm}\psi_n^{*}(\eta)Q_{nlm}^{*}(\chi_{+},\theta ,\varphi)\right); $$
\begin{equation}
\left[\hat a_{nlm}~\hat a^{+}_{n'l'm'}\right]=\delta_{nn'}\delta_{ll'}
\delta_{mm'},~~~\hat a_{nlm}|0\rangle =0,~~~|m|\leq l< n,~~~n=1,2...;
\end{equation}
$$\triangle Q_{nlm}+(n^2-1)Q_{nlm}=0; $$
\begin{equation}
Q_{nlm}=\sqrt{M_{nl}}~{P^{-l-1/2}_{n-1/2}(\cos \chi_{+})\over
\sqrt{\sin \chi_{+}}}~Y_{lm}(\theta ,\varphi),~~~
M_{nl}=\prod\limits_{p=0}^l(n^2-p^2);
\end{equation}
$$\cos \eta={1\over \cosh
H_1t_{+}},~~~a(\eta)={1\over H_1\cos\eta},~~~-{\pi\over 2}<\eta <
{\pi\over 2}; $$
\begin{equation}
\psi_n(\eta)=\sqrt{n\over 2(n^2-1)}H_1e^{-in\eta}\left(\cos\eta +
{i\sin\eta\over n}\right).
\end{equation}
Here $P_{\mu}^{\nu}$ are the Legendre functions and $\triangle$ is the
covariant Laplace operator. At the late stage of inflation
($\eta\to {\pi\over 2}$), $|\psi_n|\to H_1/\sqrt{2n(n^2-1)}$
as compared to $H_1/(2k^3)^{1/2}$ in the $K=0$ case.
So, fluctuations are slightly enhanced at low $n$ if we identify $k$
with $n$.

Gravitational waves (GW) have the same time dependence [8] with
the only difference that $n=3,4...$ for them. If the polarization tenzor
is normalized by the condition $e_{\alpha\beta}e^{\alpha\beta}=1$,
then the time dependent part $h_n=\sqrt{32\pi G}\psi_n$ [9], so
$|h_n|^2\to {16\pi GH_1^2\over n(n^2-1)}$ for $\eta \to {\pi\over 2}$
($|h_n|^2$ should be understood as the dispersion of a stochastic Gaussian
quantity). This gives the initial condition for GW at subsequent FRW
power-law stages.

To obtain the spectrum of adiabatic perturbations, one has to use
either the equation for the gravitational potential $\Phi=\Psi$ [10]:
\begin{equation}
\ddot\Phi_n+\left(H-2{\ddot\phi \over \dot\phi}\right)
\dot\Phi_n+\left({n^2-5\over a^2}+2\dot H-2{\ddot\phi \over
\dot\phi}H\right)\Phi_n=0,~~~n=3,4...,
\end{equation}
or a master equation for the generalized scalar field perturbation
$\zeta$ [11]. $\Phi_n$ is related to the Lifshitz variables $\mu_n$ and
$\lambda_n$ by the formula
\begin{equation}
\Phi_n=-{1\over 6}(\mu_n+\lambda_n)+{a\dot a \dot\lambda_n\over
2(n^2-1)}.
\end{equation}
Using any of these equations it is possible to show that $\Phi_n$
approaches the standard form at $\eta \to {\pi\over 2}$ (i.e., in the
region where both the wavelength and the radius of spatial curvature
much exceed the Hubble radius):
\begin{equation}
\Phi_n=C_n\left(1-{H\over a}\int_0^{t_{+}}adt\right),~~~
|C_n|={3H_1^3\over V_1'}{1\over \sqrt{2n(n^2-1)}}.
\end{equation}
Therefore, deviation from the flat case for adiabatic perturbations
is the same as for gravitational waves: $|C_n|\propto |h_n|$.

When calculating the CMB temperature anisotropy produced by
adiabatic perturbations and gravitational waves with these initial spectra
in a closed FRW universe, it appears that the main effect comes not so
much from the change in the initial spectrum but from an integral
term in the Sachs-Wolfe effect resulting from a deviation of a closed
matter-dominated FRW universe from a power-law expansion at recent times.
Similar to the case of a flat universe with the cosmological constant, it
results in noticeable increase of low multipoles in case of adiabatic
perturbations [12]. On the other hand, in case of $\Delta T/T$ produced by GW,
low multipoles ($l=2-4$) decrease as compared to the $K=0$ case with the
same initial amplitude while higher multipoles slightly increase. However,
the magnitude of the effect is significant for the quadrupole only ($Q$ goes
down by $15\%$ if $\Omega_m=2$) [12,13].

Since $V_1'\ll H^2_1/\sqrt{G}$, the relative ratio $T/S$ of
tensor and scalar contributions to the large-angle CMB anisotropy
is small in a closed inflationary universe. Therefore, the observational
prediction is the enhancement of low $\Delta T/T$ multipoles
relative to the dependence $C_l\equiv \langle |a_{lm}|^2\rangle
\propto (l(l+1))^{-1}$ which takes place for the flat ($n_s=1$)
spectrum in the $K=0$ case.  In particular, the $rms$ value of the
quadrupole $Q$ becomes $30\%$ larger for $\Omega_m=1.5$ and
$40\%$ larger for $\Omega_m=2.0$ if the spectrum is normalized at the $l=10$
multipole [12]. No such an enhancement is seen in the 2-year COBE data.
One may conclude that certainly $\Omega_m<2$, and probably even $\Omega_m=1.5$
can be excluded. Thus, no much place for a closed universe remains.

Note that a similar constraint can be obtained without any assumptions
about an initial perturbation spectrum, simply from the age of the
Universe $T$. In a closed matter-dominated universe,
\begin{equation}
T={2\over 3H_0}{\cal K}_T(\Omega),~~~{\cal K}_T(\Omega)={3\over
2(\Omega_m-1)}\left( {\Omega_m\over 2\sqrt{\Omega_m-1}}
\arcsin {2\sqrt{\Omega_m-1}\over \Omega_m}-1\right)
\end{equation}
(${\cal K}_T=1$ for $\Omega_m=1$).
If $H_0\geq 50$ km/s/Mpc and $T\geq 11$ Gyrs (that seems to
be the lowest value permitted by cosmic nucleosynthesis and ages
of globular clusters), then $\Omega_m\leq 2.0$. If we raise the
lower bound for $T$ to $12$ Gyrs, then $\Omega_m\leq 1.45$.

\vspace{0.5cm}
\centerline{3. Open FRW universes.}

\vspace{0.5cm}

As was discussed above, the open chart (3) covers the interior
of the future light cone of a 4-point (an event) in the de
Sitter space-time. So, one may think of an open inflationary
universe as resulting from creation of a bubble of a new de Sitter
phase in the old de Sitter phase as a result of the first order
phase transition [14,15].
As in the case $K=+1$, the duration of the second de Sitter (inflationary)
phase should be fine-tuned to about $70$ e-folds. This can be achieved by
introducing at least one additional parameter to the effective
Lagrangian describing an inflationary stage with the phase transition
during it.

We shall further consider the nearly degenerate case with practically equal
vacuum energy densities in both phases and neglect the energy density
of the bubble wall. In this approximation, the words about the
"phase transition" serve only to justify the choice of
the perturbation breaking the full de Sitter invariance
which depends on the time $t_{-}$ only. In other words, the de Sitter
symmetry breaks to the $O(2,1)$ symmetry. So, we return to the spatially
homogeneous decay of the inflationary stage investigated in [2].

Then the problem about inhomogeneous fluctuations of quantum fields
generated during this inflationary stage reduces to the quantization of a
massless scalar field $\psi$ in the chart (3). One possible consistent
quantization may be formally obtained from that in the $K=+1$ case by the
change $\chi_{+}\to i\chi_{-}$, $n\to ik$ ($k$ - real), $\eta\to {\pi\over 2}
-i\eta$. As a result, we get:
$$\hat\psi=\sum\limits_{l,m}\int_0^\infty dk(\hat a_{klm}\psi_k(\eta)
Q_{klm}(\chi_-,\theta ,\varphi) +\hat a^{+}_{klm}\psi^*_k(\eta)
Q^*_{klm}(\chi_-,\theta ,\varphi)); $$
\begin{equation}
\left[\hat a_{klm}~\hat
a^+_{klm}\right]=\delta(k-k')\delta_{ll'}\delta_{mm'},~~~a_{klm}
|\tilde 0\rangle = 0;
\end{equation}
$$\triangle Q_{klm}+(k^2+1)Q_{klm}=0; $$
\begin{equation}
Q_{klm}=\sqrt{N_{l}}~{P^{-l-1/2}_{ik-1/2}(\cosh \chi_{-})\over
\sqrt{\sinh\chi_{-}}}~Y_{lm}(\theta ,\varphi),~~~
N_l=\prod\limits_{p=0}^l(k^2+p^2);
\end{equation}
$$\sinh |\eta|={1\over \sinh H_1t_{-}},~~~a(\eta)={1\over
H_1\sinh |\eta|},~~~-\infty<\eta<0; $$
\begin{equation}
\psi_k(\eta)=\sqrt{{k\over 2(k^2+1)}}H_1e^{-ik\eta}\left(\sinh\eta-{i\cosh
\eta\over k}\right).
\end{equation}

This leads to $rms$ values of fluctuations produced at the end of the
de Sitter stage ($\eta\to 0$): $(\psi_k)_{rms}=H_1/\sqrt{2k(k^2+1)}$ [16].
However, due to the fact that $t_{-}=0$ ($\eta=-\infty$) is the
(particle) horizon and from the analogy with the quantization in the flat
space-time in the Milne metric, we know that $|\tilde 0\rangle$ is not the
correct "vacuum" state! Moreover, it can be shown that the average
energy-momentum tensor of the field $\psi$ is not regular at
$\eta\to -\infty$.

An additional complication follows from the fact that
the hypersurface $t_{-}=const$ is not the Cauchy hypersurface of the full
de Sitter space-time.
Thus, we cannot assume that perturbations are regular or square
integrable at $\chi_{-}\to \infty$.
As a result, one should either abandon the assumption that different
modes are uncorrelated (which follows from the commutator condition (17)),
or add some terms not included into the complete orthonormal set (18)
which are determined by a boundary condition at the horizon.
The de Sitter-invariant quantization of a massive scalar field
performed in [17] shows that in the massless limit one has
to add the $k^2=-1$ mode. However, some divergences remain in
the case of tensor perturbations (gravitational waves) [18], so
the question is far from being clear.

On the other hand, even the assumption of the de Sitter invariance
of the Heisenberg quantum state
is not justified in this case because the chart (3) covers only
a part of the de Sitter space-time. So, here we shall use a different
approach to calculate fluctuations of a test massless
scalar field generated during the inflationary stage in the open chart (3).
This test field may serve, e.g., as a toy model for isocurvature
perturbations.

The idea is simply to take the Green function of the massless
scalar field $\psi$ for a real inflationary stage that begins at the
moment $t=0$ in the flat chart (2) (or at the moment $t_{+}=0$
in the full chart (1) - this makes no difference as we shall see)
and to reduce it to the chart (3) using the formulas (4,5).
It is known [19-21] that this Green function is not de Sitter
invariant:
\begin{equation}
G(t_A,\vec r_A;t_B,\vec r_B)\equiv <\psi (t_A,\vec r_A)\psi (t_B,\vec
r_B)>={H_1^3\over 8\pi^2}(t_A+t_B)-{H_1^2\over 8\pi^2}\ln |z_{AB}|
+ const;
\end{equation}
$$z_{AB}=\cosh s_{AB}= \cosh (H_1(t_A-t_B))-{a_1^2\over 2}e^{H_1(t_A+t_B)}
|\vec r_A-\vec r_B|^2,~~~~K=0, $$
\begin{equation}
z_{AB}=\cosh H_1t_{-A}\cosh H_1t_{-B}-\sinh H_1t_{-A}\sinh H_1t_{-B}
\cosh \zeta_{AB},~~~K=-1,
\end{equation}
\begin{equation}
\cosh \zeta_{AB}=\cosh \chi_{-A}\cosh \chi_{-B}-\sinh \chi_{-A}\sinh
\chi_{-B}\cos \theta_{AB}
\end{equation}
where $s_{AB}$ is the geodesic distance between the 4-points A
and B , $\zeta_{AB}$ is the $3D$ geodesic distance between the 3-points
($\chi_{-A},\theta_A,\varphi_A$) and ($\chi_{-B},\theta_B,\varphi_B$) and
$\theta_{AB}$ is the angle between unit 3-vectors with the angular
directions ($\theta_A,\varphi_A$) and ($\theta_B,\varphi_B$).

We assume that the first phase of inflation before the bubble formation
is long: $H_1t_c\gg 1$. Then in the limit $H_1t_{-}\gg 1$ corresponding to the
end of the second inflationary phase, the coordinate transformation from the
$K=0$ case to $K=-1$ case reduces to:
\begin{eqnarray}
r=r_c\tanh{\chi_{-}\over 2},~~~r_c=2(a_1H_1e^{H_1t_c})^{-1}, \nonumber \\
H_1(t-t_c)=H_1t_{-}+\ln{1+\cosh \chi_{-}\over 2}.
\end{eqnarray}
Substituting (23) into (20), we obtain the answer:
\begin{equation}
G(t_{-A},\vec r_A;t_{-B},\vec r_{B})={H^2\over 8\pi^2} \ln{(1+\cosh
\zeta_{AC})(1+\cosh \zeta_{BC})\over \cosh \zeta_{AB}-1}+const ,
\end{equation}
where $C$ is the point $\chi_- =0$ - the "center" of the open
universe. The Green function (24) is stationary, this means that
fluctuations are time-independent outside the horizon as they should be.
On the other hand (and this is a new and unexpected
result), it is neither translationally invariant, nor isotropic.
So, we get a spontaneous breakdown of homogeneity and isotropy even in the
statistical sence.

Since we didn't specify the nature of the field $\psi$, it may
appear that $\psi$ itself is not observable, and only its
differences can be measured (e.g., this takes place in case of the
gravitational potential $\Phi$). Then let us introduce the observable
quantity $\tilde \psi=\psi(\vec r)-\psi(\vec r_O)$ where $\vec r_O$
is the observer (i.e., our) location.
The correlation function of $\tilde \psi$ has the form:
\begin{equation}
\tilde G\equiv <\tilde \psi(\vec r_A)\tilde \psi(\vec r_B)>=
{H^2\over 8\pi^2}\ln{(\cosh \zeta_{AO}-1)(\cosh \zeta_{BO}-1)\over
\cosh \zeta_{AB}-1}+const.
\end{equation}
It is isotropic with respect to the observer, but observer-dependent
clearly. Suppose that $\psi$ directly produces $\Delta T/T$ at the
last scattering surface. Then, shifting the center of coordinates to the
observer location and taking $\chi_{-A}=\chi_{-B}=\chi_{hor}$, we get
the correlation function $G(\theta)$:
\begin{equation}
G(\theta)={H^2\over 8\pi^2}\left(\ln{1\over 1-\cos \theta_{AB}} +
const\right).
\end{equation}
This $G(\theta)$ just coinsides with that produced by adiabatic
perturbations with the flat ($n_s=1$) spectrum in the $K=0$ case, the
corresponding dispersion of multipoles is $C_l\propto (l(l+1))^{-1}$.
Thus, in open inflationary models we expect no damping of low multipoles with
$l<\Omega_m^{-1}$, contrary to what happens in genuine open FRW models [22].
Actually, when the integral term in the Sachs-Wolfe effect is taken into
account, low multipoles will be additionally amplified. Therefore, the total
expected effect for $\Delta T/T$ multipoles in an open inflationary
universe is qualitatively the same as in a closed inflationary universe:
enhancement of low multipoles above the law $C_l\propto (l(l+1))^{-1}$.
From the absence of such an enhancement in the COBE data,
restrictions on $\Omega_m$ can be obtained which we will not discuss here.
Note also that the anisotropy of the correlation function
$G$ (24) may become observable in models where $H$ is not
constant during the first phase of inflation. However, this effect
will be proportional to ($n_s-1$) and rather small.

So, the final conclusion is that there are still some place for
inflationary models with $\Omega_m\not= 1$ though we don't see any
specific observational effect (e.g., enhancement of low multipoles in
$\Delta T/T$ above the level predicted by standard inflation) which would
push us to accept these models.

This research was partially supported by the Russian research project
"Cosmomicrophysics" through Cosmion and by the INTAS grant 93-493.

\vspace{0.5cm}

\centerline{\bf REFERENCES}

\vspace{0.5cm}

\noindent
1. A.A. Starobinsky, this volume.          \\
2. A.A. Starobinsky. Phys. Lett., {\bf B91}, 99, 1980. \\
3. E.P. Tryon. Nature, {\bf 246}, 396, 1973.   \\
4. P.I. Fomin. Preprint ITF-73-137, Kiev, 1973;
   Dokl. Akad. Nauk. Ukr. SSR, {\bf A9}, 831, 1975. \\
5. Ya.B. Zeldovich. Sov. Astron. Lett., {\bf 7}, 322, 1981. \\
6. L.P. Grishchuk, Ya.B. Zeldovich. In: {\it Quantum structure
   of space-time}, eds. M. Duff and C.I. Isham, Cambridge Univ. Press,
   p 409, 1982.                                               \\
7. N.A. Chernikov, E.A. Tagirov. Ann. Inst. Henri Poincar\'e,
   {\bf 9A}, 109, 1968.                                         \\
8. E.M. Lifshitz. Zh. Eksp. Teor. Fiz., {\bf 16}, 587, 1946.      \\
9. A.A. Starobinsky. JETP Lett., {\bf 30}, 682, 1979.               \\
10. J.J.Halliwell, S.W. Hawking. Phys. Rev. D, {\bf 31}, 1777, 1985. \\
11. K.K. Olkhin, this volume.       \\
12. T. Souradeep, A.A. Starobinsky, in preparation.   \\
13. B. Allen, R. Caldwell, S. Koranda. Preprint astro-ph/9410024, 1994. \\
14. J.R. Gott, III. Nature, {\bf 295}, 304, 1982.          \\
15. M. Bucher, A.S. Goldhaber, N. Turok. Preprint hep-ph/9411206, 1994. \\
16. D.Lyth, E.D. Stewart. Phys. Lett., {\bf B252}, 336, 1990.   \\
17. M. Sasaki, T. Tanaka, K. Yamamoto. Phys. Rev. D, {\bf 51}, 2979, 1995. \\
18. B. Allen, private communication.   \\
19. A.D. Linde. Phys. Lett., {\bf B116}, 335, 1982.      \\
20. A.A. Starobinsky. Phys. Lett., {\bf B117}, 175, 1982.    \\
21. A. Vilenkin, L.H. Ford. Phys. Rev. D, {\bf 26}, 1231, 1982. \\
22. M.L. Wilson. Astroph. J., {\bf 273}, 2, 1983.

\vfill
\end{document}